\documentclass{article}

\usepackage{amsfonts}
\usepackage{amsmath}
\usepackage{amssymb}
\usepackage{graphicx}
\usepackage{geometry}
\usepackage{hyperref}
\usepackage{cite}
\providecommand{\U}[1]{\protect\rule{.1in}{.1in}}

\begin{document}

\title{2D Relativistic Oscillators with a Uniform Magnetic Field in Anti--deSitter Space}
\author{{Lakhdar Sek}$^{1}$ \and {Mokhtar Falek}$^{1}$ \and {Mustafa Moumni}$^{1,2,a}$ \\
$^{1}$Laboratory of Photonic Physics and Nano-Materials (LPPNNM)\\
Department of Matter Sciences, University of Biskra, ALGERIA \\
$^{2}$Laboratory of Radiation and their Interactions with Matter (PRIMALAB)\\
Department of Physics, University of Batna1, ALGERIA\\
$^{a}$correspondant author m.moumni@univ-biskra.dz}
\maketitle

\begin{abstract}

We study analytically the two-dimensional deformed bosonic oscillator equation for charged particles (both spin 0 and spin 1 particles) subject to the effect of an uniform magnetic field. We consider the presence of a minimal uncertainty in momentum caused by the Anti--de Sitter model and we use the Nikiforov--Uvarov (NU) method to solve the system. The exact energy eigenvalues and the corresponding wave functions are analytically obtained for both Klein Gordon and scalar Duffin-Kemmer-Petiau (DKP) cases. For spin 1 DKP case, we deduce the behavior of the DKP equation and write the non-relativistic energies and we show the fundamental role of the spin in this case. Finally, we study the thermodynamic properties of the system.

PACS: 03.65.Ge, 03.65.Pm.

\end{abstract}
\maketitle

\section{Introduction}

Among the various attempts to incorporate gravity in the quantum world, there is an area that has generated a lot of interest; it is quantum field theory in curved space through generalizations of the Heisenberg algebra, such us the extended uncertainty principle (EUP). The purpose of this extended principle is to take into account the quantum fluctuations of the gravitational field in order to incorporate gravity into quantum mechanics. One of the consequences of this unification is the existence of a minimum length scale of the Planck order \cite{S. Mignemi}. We can link this minimum length to a modification of the standard Heisenberg algebra by adding small corrections to the canonical commutation relations and thus, we shift their standard algebra; we quote here the work of Mignemi \cite{S. Mignemi}, who showed that the Heisenberg relations are modified in the (anti--)deSitter space by adding corrections that are proportional to the cosmological constant. This modifications were also motivated by Doubly Special Relativity (DSR) \cite{G. Amelino-Camelia, G. Amelino-Camelia1}, string theory \cite{S. Capozziello}, non-commutative geometry \cite{M.R. Douglas}, black hole physics \cite{F.
Scardigli, F. Scardigli and R. Casadio} and even by the effects of Newton's gravity on quantum systems \cite{V.E. Kuzmichev}.

In the past few years, a large amount of research works has been devoted to the study of the relativistic quantum mechanics with the EUP \cite{B. Hamil, W.S. Chung, M. Hadj Moussa}. Some problems have also been solved in non-relativistic quantum mechanics; this was done despite the fact that we can not derive any non-relativistic covariant Schr\"{o}dinger--like equation from the covariant Klein--Fock--Gordon equation in the traditional field theory method of deSitter (dS) and anti--deSitter (AdS) models. Here, we use the EUP formulation to write the Schr\"{o}dinger equation version for both dS and AdS cases \cite{M. Merad, S. Ghosh, M. Falek}.

On the other hand, the studies of the thermal properties of some quantum systems were carried out recently within the framework of ordinary quantum mechanics and its deformed version by several authors. We cite here the thermodynamic properties of the ordinary Dirac oscillator which has generated considerable interest because of its description of the quark-gluon plasma model \cite{Moreno, Pacheco, Ravndal}. We also mention the thermodynamic quantities for a linear potential for Klein--Gordon (KG) and Dirac equations \cite{Arda} and for a one-dimensional Schr\"{o}dinger equation with a harmonic oscillator plus an inverse square potential \cite{Dong}. Some aspects of the bosonic oscillator has also been studied in a thermal bath in deformed quantum mechanics \cite{Nouicer, Wang, Falek1}.

Besides these systems, there is a growing interest in two-dimensional (2D) systems describing the dynamics of a charged particle confined with a strong and uniform external magnetic field. This interest is due to their various applications in different fields of matter physics and chemistry \cite{Eshghi1, Eshghi2, Kryuchkov}, semiconductor structures \cite{Weishbuch}, chemical physics \cite{Baura} and molecular vibrational and rotational spectroscopy \cite{Arda2}. Therefore, a lot of works were devoted to the study of this kind of problems in both ordinary and deformed quantum mechanics. We cite here the effects of external fields on 2D systems such as the Schr\"{o}dinger oscillator \cite{Rebane}, the KG particle under a pseudo-harmonic oscillator interaction \cite{Ikhdaira}, the Dirac equation \cite{Menculini}, the Duffin--Kemmer--Petiau (DKP) equation in cosmic string background \cite{Darroodi}, the Schr\"{o}dinger equation with minimal length in non-commutative phase space \cite{Hassanabadi1}, the KG oscillator with the presence of a minimal length in the noncommutative space \cite{Shu-Rui}, the Dirac oscillator in deformed space \cite{Boumali}, the relativistic oscillators in a noncommutative space \cite{Mirza} and the bosonic oscillator equation with the Snyder-de Sitter model \cite{Falek2}.

In this paper, we study analytically in 2D spaces, both KG and scalar DKP equations in the position space representation for deformed quantum mechanics with EUP. We consider an oscillator-like interaction and an external uniform magnetic field for this system. We also determine the thermodynamic properties of our model.

The paper is organized as follows: In Sec.2, we provide an analysis of the AdS model while in Sec.3, we introduce the Nikiforov--Uvarov (NU) method used to solve the equations of our system. We expose in Sec.4 the explicit calculations of both eigenfunctions and eigenvalues of the deformed 2D KG oscillator under a uniform magnetic field with AdS algebra. The same calculations are done in Sec.5, for the DKP version of this system. We investigate the thermodynamic properties of our model in the high temperatures regime in Sec.6 and finally, we give the concluding remarks in Sec.7.

\section{Review of the deformed quantum mechanics relation}

The deformed Heisenberg algebra leading to\ the EUP in AdS model is defined in 3D spaces by the following commutation relations \cite{S. Mignemi1, M.M. Stetsko}:
\begin{equation}
\left[  X_{i},X_{j}\right]  =0\text{ },\text{ }\left[  P_{i},P_{j}\right]
=-i\hbar\lambda\epsilon_{ijk}L_{k}\text{ },\text{ }\left[  X_{i},P_{j}\right]
=i\hbar\left(  \delta_{ij}-\lambda X_{i}X_{j}\right)  \label{1}
\end{equation}
where $\lambda$ is a small positive deformation parameter. In the sense of quantum gravity, this $\lambda$ parameter is calculated as the fundamental constant associated with the scale factor of the expanding universe and it is proportional to the cosmological constant $\Gamma=-3\lambda=-3a^{-2}$\ where $a$\ is the AdS radius \cite{B. Bolen}.

$L_{k}$ are the usual angular momentum components and they are expressed as follows:
\begin{equation}
L_{k}=\epsilon_{ijk}X_{i}P_{i} \label{2}
\end{equation}
These components follow the usual momentum algebra:
\begin{equation}
\left[  L_{i},P_{j}\right]  =i\hbar\varepsilon_{ijk}P_{k}\text{ },\text{
}[L_{i},X_{j}]=i\hbar\varepsilon_{ijk}X_{k}\text{ },[L_{i},L_{j}
]=i\hbar\varepsilon_{ijk}L_{k} \label{3}
\end{equation}
The AdS deformed algebra $\ref{1}$ gives rise to modified Heisenberg uncertainty relations:
\begin{equation}
\Delta X_{i}\Delta P_{i}\text{ }\geq\frac{\hbar}{2}\left(  1+\lambda\left(
\Delta X_{i}\right)  ^{2}\right)  \label{4}
\end{equation}
where we have chosen the states for which $\langle X_{i}\rangle=0$.

It also generates a minimum uncertainty in momentum. For simplicity, if we assume isotropic uncertainties $X_{i}=X$, we get:
\begin{equation}
(\Delta P_{i})_{min}=\hbar\sqrt{\lambda} \label{5}
\end{equation}
So the noncommutative operators $X_{i}$ and $P_{i}$ satisfy the AdS algebra $\ref{1}$ with the rescaled uncertainty relations in position space $\ref{4}$. In what follows, we represent these operators as functions of the usual $x_{i}$ and $p_{i}$\ operators fulfilling the ordinary canonical commutation relations in position space; this is done with the following transformations:
\begin{equation}
X_{i}=\frac{x_{i}}{\sqrt{1-\lambda r^{2}}} \label{6}
\end{equation}
\begin{equation}
P_{i}=-i\hbar\sqrt{1-\lambda r^{2}}\partial x_{i} \label{7}
\end{equation}
Here the variable $r$ vary in the domain $\left] -1/\sqrt{\lambda} , 1/\sqrt{\lambda}\right[$.

\section{Nikiforov-Uvarov method}

Primarily, the Nikiforov-Uvarov (NU) approach was built on the hypergeometric differential equation. The formulas used in the NU method reduce the second order differential equations to the hypergeometric kind with a suitable coordinate transformation (note that $s\equiv s(x)$ and the primes denote the derivatives):
\begin{equation}
\psi^{^{\prime\prime}}(s)+\frac{\widetilde{\tau}(s)}{\sigma(s)}\psi^{^{\prime
}}(s)+\frac{\widetilde{\sigma}(s)}{\sigma^{2}(s)}\psi(s)=0 \label{8}
\end{equation}
where $\sigma(s)$ and $\widetilde{\sigma}(s)$ are at most second degree polynomials while the degree of the polynomial $\widetilde{\tau}$ $(s)$ is strictly less than 2 \cite{H. Egrifes}\cite{A.F. Nikiforov}. If we use the following factorization:
\begin{equation}
\psi(s)=\phi(s)y(s) \label{9}
\end{equation}
eq.$\ref{8}$ becomes \cite{A.F. Nikiforov}:
\begin{equation}
\sigma(s)y^{^{\prime\prime}}(s)+\tau(s)y^{^{\prime}}(s)+\Lambda y(s)=0 \label{10}
\end{equation}
where:
\begin{equation}
\pi(s)=\sigma(s)\frac{d}{ds}(\ln\phi(s))\text{ and }\tau(s)=\widetilde{\tau }(s)+2\pi(s) \label{11}
\end{equation}
and $\Lambda$ is defined by:
\begin{equation}
\Lambda_{n}+n\tau^{^{\prime}}+\frac{n(n+1)}{2}\sigma^{^{\prime\prime}}=0\text{and }n=0,1,2,... \label{12}
\end{equation}
The energy eigenvalues of the system are determined from the equation above. To find their expressions, we have to evaluate $\pi(s)$ and $\Lambda$ first by identifying:
\begin{equation}
k=\Lambda-\pi^{^{\prime}}(s) \label{13}
\end{equation}
We get the solution of the quadratic equation for $\pi(s)$ which is a polynomial of $s$:
\begin{equation}
\pi(s)=\left(  \frac{\sigma^{^{\prime}}-\widetilde{\tau}}{2}\right)  \pm
\sqrt{\left(  \frac{\sigma^{^{\prime}}-\widetilde{\tau}}{2}\right)
^{2}-\widetilde{\sigma}+\sigma k} \label{14}
\end{equation}

It must be noted that in the calculation of $\pi(s)$, the determination of $k$ is the critical point and it is achieved by stating that the expression under the square root in $\ref{14}$ must be a polynomial square; this gives us a quadratic general equation for $k$. We use $\ref{11}$ and the Rodrigues relation to evaluate the polynomial solutions $y_{n}(s)$:
\begin{equation}
y_{n}(s)=\frac{C_{n}}{\rho(s)}\frac{d^{n}}{ds^{n}}\left[  \sigma^{n}(s)\rho(s)\right]  \label{15}
\end{equation}
where $C_{n}$ are constants used to normalize the solutions. The weight function $\rho(s)$ meets the following relation:
\begin{equation}
\frac{d}{ds}\left[  \sigma(s)\rho(s)\right]  =\tau(s)\rho(s) \label{16}
\end{equation}
This last equation refers to the classical orthogonal polynomials and we write the orthogonality relations for the polynomial solutions as follows:
\begin{equation}
\int\limits_{a}^{b}y_{n}(s)y_{m}(s)\rho(s)ds=0\text{ if }m\neq n \label{17}
\end{equation}

\section{KG oscillator in a magnetic field}

We consider the stationary relativistic equation describing a KG harmonic oscillator with a constant magnetic field in a 2D space \cite{W. Greiner}:
\begin{equation}
c^{2}\left(  \mathbf{p}-\frac{e}{c}\mathbf{A}+im\omega\mathbf{r}\right)
\cdot\left(  \mathbf{p}-\frac{e}{c}\mathbf{A}-im\omega\mathbf{r}\right)
\Psi\left(  \mathbf{r}\right)  =\left(  E^{2}-m^{2}c^{4}\right)  \Psi\left(
\mathbf{r}\right)  \label{18}
\end{equation}
We choose the $z$-axis as the magnetic field direction and use the Coulomb gauge:
\begin{equation}
\mathbf{A=}\frac{1}{2}\mathbf{B}\times\mathbf{r=}\frac{B}{2}\left(-y,x,0\right)  \label{19}
\end{equation}
Here $B$ represents the intensity of the magnetic field.

We use the AdS algebra definition (eqs.$\ref{6}$ and $\ref{7}$) to rewrite this equation in the deformed momentum space:
\begin{equation}
c^{2}(\mathbf{p}^{+}\cdot\mathbf{p}^{-})\Psi\left(  \mathbf{r}\right)
=(E^{2}-m^{2}c^{4})\Psi\left(  \mathbf{r}\right)  \label{20}
\end{equation}
with the following definitions:
\begin{equation}
\mathbf{p}^{\pm}=\mathbf{p}^{^{\prime}}\pm im\omega\frac{\mathbf{r}}
{\sqrt{1-\lambda r^{2}}},\mathbf{p}^{^{\prime}}=\sqrt{1-\lambda r^{2}
}\mathbf{p}-\frac{e}{c}\mathbf{B}\times\frac{\mathbf{r}}{\sqrt{1-\lambda r^{2}}}\label{21}
\end{equation}
Following a straightforward calculation, we get the following equation:
\begin{equation}
\left[  (1-\lambda r^{2})p^{2}+\eta\frac{r^{2}}{1-\lambda r^{2}}+i\hbar
\lambda(\mathbf{r}\cdot\mathbf{p})-\frac{eB}{c}L_{z}-\varepsilon\right]
\Psi\left(  \mathbf{r}\right)  =0\label{22}
\end{equation}
with the parameters:
\begin{equation}
\eta=m^{2}\omega^{2}+\frac{e^{2}B^{2}}{4c^{2}}-\lambda\hbar m\omega\text{ and }\varepsilon=\frac{(E^{2}-m^{2}c^{4})}{c^{2}}+2m\omega\hbar\label{23}
\end{equation}
To get the exact solution of eq.$\ref{22}$, we use the polar coordinates in momentum space $\left(  r,\varphi\right)$ and write the solutions in a separate form containing the azimuthal quantum number $l$:
\begin{equation}
\Psi\left(  r\mathbf{,}\varphi\right)  =\exp(il\varphi )R(r),l=0,1,2,...\label{24}
\end{equation}
So, eq.$\ref{22}$ transforms to the following expression:
\begin{equation}
\left[  \left(  \sqrt{1-\lambda r^{2}}\frac{d}{dr}\right)  ^{2}+\frac
{1-\lambda r^{2}}{r}\frac{d}{dr}-\frac{l^{2}\left(  1-\lambda r^{2}\right)
}{r^{2}}-\frac{\eta r^{2}}{\hbar^{2}\left(  1-\lambda r^{2}\right)  }
+\epsilon\right]  R(r)=0\label{25}
\end{equation}
with:
\begin{equation}
\text{ }\epsilon\mathbf{=}\frac{\varepsilon}{\hbar^{2}}+\frac{eBl}{c\hbar }\label{26}
\end{equation}
Now to solve eq.$\ref{25}$, we use the following transformation:
\begin{equation}
R(\rho)=\rho^{\mu}g\left(  \rho\right)  \text{ and }\rho=\sqrt{1-\lambda r^{2}}\label{27}
\end{equation}
By doing this, eq.$\ref{25}$ gives us:
\begin{equation}
\left[  \left(  1-\rho^{2}\right)  \dfrac{d^{2}}{d\rho^{2}}+2\left(  \frac
{\mu}{\rho}-\left(  \mu+1\right)  \rho\right)  \dfrac{d}{d\rho}-\frac
{l^{2}\rho^{2}}{\left(  1-\rho^{2}\right)  }+\frac{\epsilon}{\lambda}
-2\mu\right]  g\left(  \rho\right)  =0.\label{28}
\end{equation}
where we have chosen the free parameter $\mu$ so that it satisfies the relation:
\begin{equation}
\mu\left(  \mu-1\right)  -\frac{\eta}{\hbar^{2}\lambda^{2}}=0\label{29}
\end{equation}
The solutions of this equation are given by:
\begin{equation}
\mu=\frac{1}{2}\pm\sqrt{\frac{1}{4}+\frac{\eta}{\hbar^{2}\lambda^{2}}}\label{30}
\end{equation}
From eq.$\ref{28}$, we see that $g\left(  \rho\right)  $ should be nonsingular at $\rho=\pm1$ and the same is true for $R(\rho)$ from eq.$\ref{27}$; thus the accepted\ value of $\mu\ $ is:
\begin{equation}
\mu=\frac{1}{2}+\sqrt{\frac{1}{4}+\frac{\eta}{\hbar^{2}\lambda^{2}}}\label{31}
\end{equation}
We note that eq.$\ref{28}$ possesses three singular points $\rho=0,1,-1$ and to reduce it to a class of known differential equation with a polynomial solution, we use another change of variable $s=2\rho^{2}-1$:
\begin{equation}
\left[  \frac{d^{2}}{ds^{2}}+\frac{\left(  \mu-\frac{1}{2}\right)  -\left(
\mu+\frac{3}{2}\right)  s}{1-s^{2}}\frac{d}{ds}-\frac{\left(  l^{2}
+\epsilon\right)  s^{2}+2l^{2}s-\left(  \epsilon\mathbf{-}l^{2}\right)
}{4\left(  1-s^{2}\right)  ^{2}}\right]  g\left(  s\right)  =0\label{32}
\end{equation}
Comparing eq.$\ref{32}$ with eq.$\ref{8}$ allows us to use the NU method with polynomials given by:
\begin{equation}
\sigma\left(  s\right)  =1-s^{2},\widetilde{\sigma}\left(  s\right)  =\frac
{1}{4}\left[  -\left(  l^{2}+\epsilon\right)  s^{2}-2l^{2}s+\left(
\epsilon\mathbf{-}l^{2}\right)  \right]  \text{ \& }\widetilde{\tau}\left(
s\right)  =\left(  \mu-\frac{1}{2}\right)  -\left(  \mu+\frac{3}{2}\right)
s\label{33}
\end{equation}
We replace them in eq.$\ref{14}$ to get:
\begin{align}
\pi(s) &  =\frac{\left(  \mu-\frac{1}{2}\right)  \left(  s-1\right)  }{2}
\pm\left[  \frac{1}{2}\left(  \left(  \mu-\frac{1}{2}\right)  ^{2}
+l^{2}-\left(  4k-\epsilon\right)  \right)  s^{2}-\right.  \nonumber\\
&  \left.  \left(  \left(  \mu-\frac{1}{2}\right)  ^{2}-l^{2}\right)
s+\frac{1}{2}\left(  \left(  \mu-\frac{1}{2}\right)  ^{2}+l^{2}+\left(
4k-\epsilon\right)  \right)  \right]  ^{\frac{1}{2}}\label{34}
\end{align}
The parameter $k$ is determined as mentioned in the precedent section and we get two values:
\begin{equation}
k_{1}=\frac{\epsilon}{4}\mathbf{+}\frac{l}{2}\left(  \mu-\frac{1}{2}\right)
\text{ and }k_{2}=\frac{\epsilon}{4}\mathbf{-}\frac{l}{2}\left(  \mu-\frac
{1}{2}\right)  \label{35}
\end{equation}
For $\pi(s)$, we obtain the following solutions:
\begin{equation}
\pi(s)=\left\{
\begin{array}
[c]{c}
\pi_{1,3}=\frac{1}{2}\left[  \left(  2\mu\mp l+1\right)  s-\left(  2\mu\pm
l-1\right)  \right]  \\
\pi_{2,4}=\pm\frac{l}{2}\left(  s+1\right)
\end{array}
\right.  \label{36}
\end{equation}
where $\pi_{1}$ and $\pi_{2}$ are related to $k_{1}$ while $\pi_{3}$ and $\pi_{4}$ are linked to $k_{2}$.

In our case, the relevant solution is the proper value $\pi_{4}$, so that we have:
\begin{equation}
\tau(s)=-\left(  \mu+\frac{3}{2}+l\right)  s+\left(  \mu-\frac{1}{2}-l\right) \label{37}
\end{equation}
From eqs.$\ref{12}$ and $\ref{13}$, we obtain:
\begin{equation}
\Lambda_{n}=k_{2}-\frac{l}{2}=n\left(  n+\mu+l+\frac{1}{2}\right) ,n=0,1,2,... \label{38}
\end{equation}
Hence, we found the expressions of the energy eigenvalues as:
\begin{align}
E_{n,l}  &  =\pm mc^{2}\left[  1-\frac{2\omega\hbar}{mc^{2}}+\frac{2\hbar
}{mc^{2}}\left\{  \left(  2n+l+1\right)  \sqrt{\left(  \omega-\frac
{\lambda\hbar}{2m}\right)  ^{2}+\widetilde{\omega}^{2}}+\right.  \right.
\nonumber\\
&  \left.  \left.  \frac{\lambda\hbar}{2m}\left(  4n\left(  n+l+1\right)
+2l+1\right)  -\widetilde{\omega}l\right\}  \right]  ^{\frac{1}{2}} \label{39}
\end{align}
where we have used the definition $\widetilde{\omega}=eB/2mc=\omega_{c}/2c$ with $\omega_{c}$ the cyclotron frequency.

We remark that the energy spectrum of our system contains two corrections associated with the deformation; a first one that comes with the oscillator term and an additional one which increases with the deformation parameter $\lambda$. The overall effect of the deformation on the spectrum is to increase the energy levels in proportion to the value of $\lambda$. Here it should be noted that, according to the $n^{2}$ dependence of the energies, which corresponds to a confinement at the high energy area, our result is equivalent to the energy of a spinless relativistic quantum particle in a square well potential; in our case, the boundaries of the well are placed at $\pm\pi/2\sqrt{\lambda}$.

We can test the shape of the energy spectrum as follows: It corresponds to same spectrum of the deformed 2D KG oscillator under a uniform magnetic field with Snyder-de Sitter algebra if we ignore the effect of Snyder algebra ($\alpha_{2}=0$ and $\alpha_{1}=\lambda$)\cite{Falek2} and, if we study the limit $\lambda\rightarrow0$, we obtain the exact result of the 2D KG oscillator under a magnetic field without deformation. We also note that the result is strictly consistent with the usual KG oscillator when both the deformed parameter and the magnetic field are absent (i.e. $\lambda=B=0$) \cite{Xiao}.

Now let us find the corresponding eigenfunctions. Taking the expression of $\pi_{4}\left(  s\right) $ from eq.$\ref{36}$, the $\phi\left(  s\right)  $ part is defined from the relation $\ref{11}$ as follows:
\begin{equation}
\phi\left(  s\right)  =\left(  1-s\right)  ^{l/2} \label{40}
\end{equation}
and according to the form of $\sigma\left(  s\right)  $ (eq.$\ref{33}$), the Rodrigues relation $\ref{15}$ gives us the $y\left(  s\right)  $ part:
\begin{equation}
y_{n}\left(  s\right)  =\frac{C_{n}}{\rho\left(  s\right)  }\frac{d^{n}
}{ds^{n}}\left[  \left(  1-s^{2}\right)  ^{n}\rho\left(  s\right)  \right]
\label{41}
\end{equation}
where $\rho\left(  s\right)  =\left(  1+s\right)  ^{\left(  \mu-\frac{1}
{2}\right)  }\left(  1-s\right)  ^{l}$. We see that eq.$\ref{41}$ stands for the Jacobi polynomials:
\begin{equation}
y_{n}\left(  s\right)  \equiv P_{n}^{\left(  l,\mu-1/2\right)  }\left(
s\right)  \label{42}
\end{equation}
Hence, $f(s)$ can be written in the following form:
\begin{equation}
f(s)=C_{n}\left(  1-s\right)  ^{\frac{l}{2}}P_{n}^{\left(  l,\mu-1/2\right)
}\left(  s\right)  \label{43}
\end{equation}
In terms of the variables $r$ and $\varphi$, we can now write the general form of the wave function $\Psi$:
\begin{equation}
\Psi\left(  r\mathbf{,}\varphi\right)  =C_{n}2^{\frac{l}{2}}e^{il\varphi
}\left(  1-\lambda r^{2}\right)  ^{\frac{\mu}{2}}\left(  \lambda r^{2}\right)
^{\frac{l}{2}}P_{n}^{\left(  l,\mu-1/2\right)  }\left(  1-2\lambda
r^{2}\right)  \label{44}
\end{equation}
where\ $\mu$ is given in eq.$\ref{31}$ and the constant $C_{n}$ is obtained using the normalization condition in the space of the radial wave function \cite{M.M. Stetsko}:
\begin{equation}
\int_{0}^{1/\sqrt{\lambda}}\frac{2^{l+1}\pi rdr}{\left(  1-\lambda r^{2}\right)  ^{1/2}}R^{\ast}(r)R(r)=1 \label{45}
\end{equation}
and the orthogonality relation of the Jacobi polynomials \cite{I. S. Gradshteyn}:
\begin{equation}
\int_{-1}^{1}dy(1-y)^{a}(1+y)^{b}\left[  P_{n}^{(a,b)}(y)\right]  ^{2}
=\frac{2^{a+b+1}\Gamma(a+n+1)\Gamma(b+n+1)}{n!(a+b+1+2n)\Gamma\left(
a+b+n+1\right)  } \label{46}
\end{equation}
So we get the normalization constant value:
\begin{equation}
C_{n}=\sqrt{\frac{\lambda}{2^{l}\pi}\frac{n!\left(  2n+\mu+l+\frac{1}
{2}\right)  \Gamma\left(  n+\mu+l+\frac{1}{2}\right)  }{\Gamma\left(
n+\mu+\frac{1}{2}\right)  \Gamma\left(  n+l+1\right)  }} \label{47}
\end{equation}
Let us now check these solutions by studying the limits $\lambda\rightarrow0$ of the expression $\ref{44}$. We use the following relations \cite{I. S. Gradshteyn}:
\begin{equation}
\lim_{\mu\rightarrow\infty}P_{n}^{\left(  \alpha,\beta\right)  }\left(
1-\frac{2x}{\mu}\right)  =L_{n}^{\alpha}\left(  x\right)  \text{ and }
\lim_{\nu\rightarrow\infty}\frac{\Gamma\left(  v+a\right)  }{\Gamma\left(
v\right)  }\nu^{-a}=1 \label{48}
\end{equation}
and we limit ourselves to the first order of $\lambda$ to write $\left(1-\lambda r^{2}\right)  ^{\frac{\mu}{2}}=\exp\left( -\frac{\Omega r^{2}} {2}\right)  $ with $\Omega=\left(  \frac{m^{2}\omega^{2}}{\hbar^{2}} +\frac{e^{2}B^{2}}{4\hbar^{2}c^{2}}\right)  ^{1/2}$. We thus obtain directly the position space eigenfunction of the usual KGO (without deformation):
\begin{equation}
\Psi\left(  r\mathbf{,}\varphi\right)  =\sqrt{\frac{n!\Omega^{l+1}}{\pi
\Gamma\left(  n+l+1\right)  }}e^{il\varphi}\exp\left(  -\frac{\Omega r^{2}}
{2}\right)  r^{l}L_{n}^{l}\left(  \Omega r^{2}\right)  \label{49}
\end{equation}
An interesting characteristic of the spectrum appears when computing the energy levels spacing; indeed we find the limit:
\begin{equation}
\lim\limits_{n\rightarrow\infty}\Delta E_{n,l}=2\hbar c\sqrt{\lambda}
\label{50}
\end{equation}
This expression shows that, without the effects of the deformed algebra, the energy spectrum of the KG oscillator under a strong magnetic field becomes almost continuous for large values of $n$. In contrast, this continuous feature of the spectrum disappears and it reduces to a bound spectrum in the deformed case. This asymptotic behavior is described by eq.$\ref{50}$ where the spacing is proportional to the deformation parameter $\lambda$.

In order to get an upper bound for this $\lambda$ parameter, we use the $s-$states of the energies from eq.$\ref{39}$ and we expand it up to the first order in $\lambda$:
\begin{equation}
E_{n,0}=E_{n,0}^{\lambda=0}+\frac{\hbar^{2}c^{2}}{2E_{n,0}^{\lambda=0}}\left[
\left(  2n+1\right)  ^{2}-\frac{\left(  2n+1\right)  \omega}{\sqrt{\omega
^{2}+\widetilde{\omega}^{2}}}\right]  \lambda\label{51}
\end{equation}
with:
\begin{equation}
E_{n,0}^{\lambda=0}=\sqrt{m^{2}c^{4}-2\omega\hbar mc^{2}+2\left(  2n+1\right)
\hbar mc^{2}\sqrt{\omega^{2}+\widetilde{\omega}^{2}}} \label{52}
\end{equation}
These two relations show that the deviation of the $n$-th energy level caused by the modified commutation relations $\ref{1}$, is provided by:
\begin{equation}
\frac{\Delta E_{n,0}^{\lambda}}{\hbar\omega}=\frac{\hbar c^{2}}{2\omega
E_{n,0}^{\lambda=0}}\left[  \left(  2n+1\right)  ^{2}-\frac{\left(
2n+1\right)  \omega}{\sqrt{\omega^{2}+\widetilde{\omega}^{2}}}\right]
\lambda\label{53}
\end{equation}
We use the experimental results of the electron cyclotron motion in a Penning trap. Here, the cyclotron frequency of an electron trapped in a magnetic field of strength $B$ is $\omega_{c}=eB/m_{e}$ (without deformation), so we have $m_{e}\hbar\omega_{c}=e\hbar B=$ $10^{-52}kg^{2}m^{2}s^{-2}$ for a magnetic field of strength $B=6T$. If we assume that only the deviations of the scale of $\hbar\omega_{c}$ can be detected at the level $n=10^{10}$ and that $\Delta
E_{n}<\hbar\omega_{c}$ (no perturbation of the $n$-th energy level is observed) \cite{Chang}, we can write the following  constraint:
\begin{equation}
\lambda<3.36\times10^{-4}m^{-2} \label{54}
\end{equation}
This leads to the following upper bound of the minimal uncertainty in momentum $\Delta P_{\min}=\hbar\sqrt{\lambda}<2\times0^{-36}kgms^{-1}$; it is similar to that obtained in \cite{Falek3}.

For the non-relativistic limit, by setting $E=mc^{2}+E_{nr}$ with the assumption that $mc^{2}\gg E_{nr}$, we write the  spectrum of the non-relativistic KGO in the deformed AdS space as:
\begin{equation}
E_{nr}=\left(  2n+l+1\right)  \hbar\sqrt{\left(  \omega-\frac{\lambda\hbar
}{2m}\right)  ^{2}+\widetilde{\omega}^{2}}+\frac{\lambda\hbar^{2}}{2m}\left(
4n\left(  n+l+1\right)  +2l+1\right)  -\hbar\widetilde{\omega}l-\hbar
\omega\label{55}
\end{equation}

\section{\textbf{DKP }oscillator in a magnetic field}

The free DKP equation of massive scalar and vector particles $m$ can be written as follows\ \cite{Y. Nedjadi1, Y. Nedjadi2, Y. Nedjadi3, Y. Nedjadi4, R. Y. Duffin, L. Chetouani}:
\begin{equation}
\left[  c\mathbf{\beta\cdot p+}mc^{2}\right]  \Psi\left(  \mathbf{r},t\right)
=i\hbar\beta^{0}\partial_{0}\Psi\left(  \mathbf{r},t\right)  \label{56}
\end{equation}
where $\mathbf{\beta}$ and $\beta^{0}$ are the DKP matrices \cite{R. Y. Duffin}.

We write the 2D DKP oscillator in a same way as the Dirac oscillator \cite{M. Moshinsky} and so, we introduce the non-minimal substitution \cite{Y. Nedjadi1}:
\begin{equation}
\mathbf{p\rightarrow p}-im\omega\eta^{0}\mathbf{r} \label{57}
\end{equation}
where $\omega$ is the frequency of the oscillator and $\eta^{0}$ is defined by the relation $\eta^{0}=2\left(  \beta^{0}\right)  ^{2}-1$ (note that $\left(\eta^{0}\right)  ^{2}=1$).

Using the relation $\ref{57}$ in eq.$\ref{56}$, we get the equation of the DKP oscillator and we add a vector potential $\mathbf{A}=\frac{1}{2} \mathbf{B}\times\mathbf{r}$:
\begin{equation}
\left[  c\mathbf{\beta\cdot}\left(  \mathbf{p-}\frac{e}{c}\mathbf{A-}
im\omega\eta^{0}\mathbf{r}\right)  +mc^{2}\right]  \Psi=i\hbar\beta^{0}\left(
\frac{\partial\Psi}{\partial t}\right)  \label{58}
\end{equation}
By taking the definition of the AdS algebra ($\ref{1}$), we write the 2D deformed stationary DKP oscillator equation for the stationary solutions ($\Psi\left(  \mathbf{r},t\right)  =e^{-iEt/\hbar}\tilde{\Psi}\left(\mathbf{r}\right)  $):
\begin{equation}
\left[  c\mathbf{\beta\cdot}\left(  \sqrt{1-\lambda r^{2}}\mathbf{p}-\frac
{e}{c}\mathbf{B}\times\frac{\mathbf{r}}{\sqrt{1-\lambda r^{2}}}\mathbf{-}
im\omega\eta^{0}\frac{\mathbf{r}}{\sqrt{1-\lambda r^{2}}}\right)
+mc^{2}\right]  \tilde{\Psi}=E\beta^{0}\tilde{\Psi} \label{59}
\end{equation}

\subsection{Scalar particle case}

In the case of a scalar particle (spin $0$), the wave function is a vector with five components:
\begin{equation}
\tilde{\Psi}\left(  \mathbf{r}\right)  =\left(
\begin{array}
[c]{c}
\Phi\\
i\mathbf{\psi}
\end{array}
\right)  \text{\ with }\Phi\equiv\left(
\begin{array}
[c]{c}
\phi\\
\chi
\end{array}
\right)  \text{ and }\mathbf{\psi}\equiv\left(
\begin{array}
[c]{c}
\psi_{1}\\
\psi_{2}\\
\psi_{3}
\end{array}
\right)  \label{60}
\end{equation}
Let us substitute eq.$\ref{60}$ into eq.$\ref{59}$, we obtain the following coupled system:
\begin{align}
mc^{2}\phi &  =E\chi+ic\mathbf{p}^{+}\cdot\mathbf{\psi}\label{61}\\
mc^{2}\mathbf{\psi}  &  =ic\mathbf{p}^{-}\phi\label{62}\\
mc^{2}\chi &  =E\phi\label{63}
\end{align}
where the expressions of $\mathbf{p}^{+}$ and $\mathbf{p}^{-}$ are given in eq.$\ref{21}$.

At this stage, the system is uncoupled in favour of $\phi$ and so, it can be converted directly to the same differential equation of the 2D deformed KG case ($\ref{22}$):
\begin{equation}
\left[  \left(  1-\lambda r^{2}\right)  p^{2}+\frac{\eta r^{2}}{1-\lambda
r^{2}}+i\hbar\lambda\mathbf{r.p}-\frac{eB}{c}L_{z}-\varepsilon\right]  \phi=0 \label{64}
\end{equation}
where $\eta$ and $\varepsilon$ have the same definitions given in section 2 ($\ref{23}$).

Following the same reasoning as in the KG case and so, according to the condition $\ref{29}$, the exact solution of the scalar DKP is given by ($C^{\backprime}$ is a normalization constant and $\mu$ is given in $\ref{23}$):
\begin{equation}
\phi_{n}(r,\varphi)=C^{\backprime}2^{\frac{l}{2}}e^{il\varphi}\left(
1-\lambda r^{2}\right)  ^{\frac{\mu}{2}}\left(  \lambda r^{2}\right)
^{\frac{l}{2}}P_{n}^{\left(  l,\mu-\frac{1}{2}\right)  }\left(  1-2\lambda
r^{2}\right)  \label{65}
\end{equation}
and the corresponding energy spectrum is giving with:
\begin{align}
E_{n,l}  &  =\pm mc^{2}\left[  1-\frac{2\omega\hbar}{mc^{2}}+\frac{2\hbar
}{mc^{2}}\left\{  \left(  2n+l+1\right)  \sqrt{\left(  \omega-\frac
{\lambda\hbar}{2m}\right)  ^{2}+\widetilde{\omega}^{2}}+\right.  \right.
\nonumber\\
&  \left.  \left.  \frac{\lambda\hbar}{2m}\left(  4n\left(  n+l+1\right)
+2l+1\right)  -\widetilde{\omega}l\right\}  \right]  ^{\frac{1}{2}} \label{66}
\end{align}
The other components are simple to assess, and the final expressions of $\tilde{\Psi}\left(  \mathbf{r}\right) $ are as follows:
\[
\tilde{\Psi}_{n,l}(r,\varphi)=C^{\backprime}2^{\frac{l}{2}}e^{il\varphi
}\left(  1-\lambda r^{2}\right)  ^{\frac{\mu}{2}}\left(  \lambda r^{2}\right)
^{\frac{l}{2}}\times
\]
\begin{equation}
\left[  \left(
\begin{array}
[c]{c}
1\\
\dfrac{E}{mc^{2}}\\
M\left(  r\right) \\
N\left(  r\right) \\
0
\end{array}
\right)  P_{n}^{\left(  l,\mu-\frac{1}{2}\right)  }\left(  1-2\lambda
r^{2}\right)  -\left(
\begin{array}
[c]{c}
0\\
0\\
\Lambda\left(  r\right) \\
0\\
0
\end{array}
\right)  P_{n-1}^{\left(  l+1,\mu+\frac{1}{2}\right)  }\left(  1-2\lambda
r^{2}\right)  \right]  \label{67}
\end{equation}
with:
\begin{align}
M\left(  r\right)   &  =\frac{\hbar l}{mc}\frac{\sqrt{1-\lambda r^{2}}}
{r}+\left(  \omega-\frac{\lambda\hbar\mu}{m}\right)  \frac{r}{c\sqrt{1-\lambda
r^{2}}}\label{68}\\
N\left(  r\right)   &  =\frac{i}{mc}\left(  \frac{\hbar l\sqrt{1-\lambda
r^{2}}}{r}-\frac{eBr}{2c\sqrt{1-\lambda r^{2}}}\right) \label{69}\\
\Lambda\left(  r\right)   &  =\frac{2\lambda\hbar}{m}\left(  n+l+\mu+\frac
{1}{2}\right)  r\sqrt{1-\lambda r^{2}} \label{70}
\end{align}
and wa used the following properties of the Jacobi polynomials:
\begin{equation}
\frac{dP_{n}^{\left(  a,b\right)  }\left(  y\right)  }{dy}=\frac{1}{2}\left(
n+a+b+1\right)  P_{n-1}^{\left(  a+1,b+1\right)  }\left(  y\right)  \label{71}
\end{equation}
Before ending this section, we must calculate the normalization constant $C^{\backprime}$ using the following normalization condition:
\begin{equation}
\int\frac{rdr}{\sqrt{1-\lambda r^{2}}}\overline{\Psi}(r,\varphi)\beta^{0}
\Psi(r,\varphi)=1\text{ \ with }\overline{\Psi}=\Psi^{+}\eta^{0} \label{72}
\end{equation}
We use the components of the spinor $\Psi$ to write it as:
\begin{equation}
\int\frac{rdr}{\sqrt{1-\lambda r^{2}}}\left[  \phi^{\ast}\varphi+\varphi ^{\ast}\phi\right]  =1 \label{73}
\end{equation}
Then with the help of the Jacobi polynomial orthogonality relation ($\ref{46}$), we obtain:
\begin{equation}
C^{\backprime}=\sqrt{\frac{\lambda mc^{2}}{2^{l+1}\pi E}\frac{n!\left(
2n+\mu+l+\frac{1}{2}\right)  \Gamma\left(  n+\mu+l+\frac{1}{2}\right)
}{\Gamma\left(  n+\mu+\frac{1}{2}\right)  \Gamma\left(  n+l+1\right)  }}
\label{74}
\end{equation}
Thus we obtain the final form of the solutions $\tilde{\Psi}\left(\mathbf{r}\right)  $.

\subsection{Vector particle case}

The wave function of the spin $1$ particle is a vector with ten components noted by $\tilde{\Psi}\left(  \mathbf{r}\right)  ^{T}=(i\varphi,\mathbf{A}\left(\mathbf{r}\right),\mathbf{B}\left(\mathbf{r}\right),\mathbf{C}\left(\mathbf{r}\right))$, with $A_{i},B_{i}$ and $C_{i}$ ($i=1,2,3$) being respectively the components of the vectors $\mathbf{A}\left(\mathbf{r}\right),\mathbf{B}\left(\mathbf{r}\right)$ and $\mathbf{C}\left(\mathbf{r}\right)$. Putting this form in eq.$\ref{59}$ gives us the following system:
\begin{align}
mc^{2}\varphi &  =-c\,\mathbf{p}^{-}\cdot\mathbf{B}\label{75}\\
mc^{2}\mathbf{A}  &  =E\,\mathbf{B}-c\,\mathbf{p}^{+}\times\mathbf{C} \label{76}\\
mc^{2}\mathbf{B}  &  =E\,\mathbf{A}+c\,\mathbf{p}^{+}\varphi\label{77}\\
mc^{2}\mathbf{C}  &  =-c\,\mathbf{p}^{-}\times\mathbf{A} \label{78}
\end{align}
To decouple this system, we eliminate $\varphi$, $\mathbf{B}$ and $\mathbf{C}$ in terms of $\mathbf{A}$ and we get:
\begin{equation}
\left(  E^{2}-m^{2}c^{4}\right)  \mathbf{A}=c^{2}\mathbf{p}^{+}\left(
\mathbf{p}^{-}\cdot\mathbf{A}\right)  -c^{2}\mathbf{p}^{+}\times\left(
\mathbf{p}^{-}\times\mathbf{A}\right)  -\dfrac{1}{m^{2}}\mathbf{p}^{+}\left[
\mathbf{p}^{-}\cdot\left[  \mathbf{p}^{+}\times\left(  \mathbf{p}^{-}
\times\mathbf{A}\right)  \right]  \right]  \label{79}
\end{equation}
and we rewrite it in the following form:
\begin{equation}
\left(  E^{2}-m^{2}c^{4}\right)  \mathbf{A}=c^{2}\left[  \left(
\mathbf{p}^{+}\cdot\mathbf{p}^{-}\right)  \mathbf{A}-\left(  \mathbf{p}
^{+}\times\mathbf{p}^{-}\right)  \times\mathbf{A}\right]  -\dfrac{1}{m^{2}
}\mathbf{p}^{+}\left[  \mathbf{p}^{-}\cdot\left[  \mathbf{p}^{+}\times\left(
\mathbf{p}^{-}\times\mathbf{A}\right)  \right]  \right]  \label{80}
\end{equation}
A direct calculation of the first two terms in eq.$\ref{80}$ gives:
\begin{align}
\left(  \mathbf{p}^{+}\cdot\mathbf{p}^{-}\right)  \mathbf{A}  &  =\left[
\left(  m^{2}\omega^{2}+\frac{e^{2}B^{2}}{4c^{2}}-m\omega\hbar\lambda\right)
\frac{r^{2}}{1-\lambda r^{2}}+\left(  1-\lambda r^{2}\right)  p^{2}
+i\hbar\lambda\mathbf{r\cdot p}\right.  \mathbf{-}\nonumber\\
&  \left.  \frac{eB}{c}L_{z}-2m\omega\hbar\right]  \mathbf{A}\label{81}\\
\left(  \mathbf{p}^{+}\times\mathbf{p}^{-}\right)  \times\mathbf{A}  &
=\left[  \frac{eB}{c}\left(  \lambda-\frac{2m\omega}{\hbar}\right)
\frac{r^{2}}{1-\lambda r^{2}}+2\left(  \frac{eB}{c}+\left(  \lambda
+\frac{2m\omega}{\hbar}\right)  L_{z}\right)  \right]  S_{z}\mathbf{A}
\label{82}
\end{align}
Inserting these results into eq.$\ref{80}$, we get:
\begin{align}
\varepsilon^{\backprime}\mathbf{A}  &  =\left[  \left(  1-\lambda
r^{2}\right)  p^{2}+\left(  m^{2}\omega^{2}+\frac{e^{2}B^{2}}{4c^{2}}
-\frac{eB}{c}\left(  \lambda-\frac{2m\omega}{\hbar}\right)  S_{z}-m\omega
\hbar\lambda\right)  \frac{r^{2}}{1-\lambda r^{2}}\right.  +\nonumber\\
&  \left.  i\hbar\lambda\mathbf{r\cdot p-}\left(  \frac{eB}{c}+2\left(
\lambda+\frac{2m\omega}{\hbar}\right)  S_{z}\right)  L_{z}\right]
\mathbf{A}-\dfrac{1}{\left(  mc\right)  ^{2}}\mathbf{p}^{+}\left[
\mathbf{p}^{-}\cdot\left[  \mathbf{p}^{+}\times\left(  \mathbf{p}^{-}
\times\mathbf{A}\right)  \right]  \right]  \label{83}
\end{align}
with:
\begin{equation}
\varepsilon^{\backprime}=\dfrac{E^{2}-m^{2}c^{4}}{c^{2}}+2m\omega\hbar +\frac{2eB}{c}S_{z} \label{84}
\end{equation}
and $L_{z}$ and $S_{z}$ are the $z$-components of the orbital angular momentum and the spinor respectively.

To our knowledge, there is no analytical method to solve equations like eq.$\ref{83}$ but that does not prevent us from studying its nonrelativistic limit; this is performed by considering the last term as negligible since it is of the order $m^{-3}$.

Using this approximation, eq.$\ref{83}$ becomes similar to both eq.$\ref{22}$ and eq.$\ref{64}$ corresponding to the KG case and to the scalar DKP particle respectively; we find:
\begin{align}
\varepsilon^{\backprime}\mathbf{A}  &  =\left[  \left(  1-\lambda
r^{2}\right)  p^{2}+\left(  m^{2}\omega^{2}+\frac{e^{2}B^{2}}{4c^{2}}
-\frac{eB}{c}\left(  \lambda-\frac{2m\omega}{\hbar}\right)  S_{z}-\lambda\hbar
m\omega\right)  \frac{r^{2}}{1-\lambda r^{2}}\right.  +\nonumber\\
&  \left.  i\hbar\lambda\mathbf{r\cdot p-}\left(  \frac{eB}{c}+2\left(
\lambda+\frac{2m\omega}{\hbar}\right)  S_{z}\right)  L_{z}\right]  \mathbf{A} \label{85}
\end{align}

Comparing this equation to both eq.$\ref{22}$ and eq.$\ref{64}$, we find four additional terms. Two of them appear in the harmonic part ($\varpropto r^{2}$); the pure spin term $2m\omega S_{z}/\hbar$ and the interaction term $2m\omega eBS_{z}/\hbar c$. The other two are of spin-orbit type; the usual one $4m\omega S_{z}L_{z}/\hbar$ and the additional $2\lambda S_{z}L_{z}$ coming from the influence of the space deformation. As it should be, all these terms are due to the presence of the spin in this vectorial case.

We follow the same procedure used in the precedent sections, to obtain the energies and we get:
\begin{align}
E^{nr}  &  =\left(  2n+l+1\right)  \hbar\sqrt{\left(  \omega-\frac
{\lambda\hbar}{2m}\right)  ^{2}+\widetilde{\omega}^{2}-\frac{4}{\hbar
}\widetilde{\omega}\left(  \omega-\frac{\lambda\hbar}{2m}\right)  S_{z}
}-\left(  \hbar\omega+\widetilde{\omega}S_{z}\right)  +\nonumber\\
&  \frac{\lambda\hbar^{2}}{2m}\left(  4n\left(  n+l+1\right)  +2l+1\right)
-\left[  \widetilde{\omega}+\frac{2}{\hbar}\left(  \omega+\frac{\lambda\hbar
}{2m}\right)  S_{z}\right]  l \label{86}
\end{align}
The zero value of $S_{z}$ ($m_{s}=0$) gives us the same spectrum of the scalar particle (eq.$\ref{66}$). So, when considering the nonzero eigenvalues of $S_{z}$ $\left(  m_{s}=\pm1\right)  $, we write the final expression of the non--relativistic energy spectrum as follows:
\begin{align}
E^{nr}  &  =\left(  2n+l+1\right)  \hbar\sqrt{\left(  \omega-\frac
{\lambda\hbar}{2m}\right)  ^{2}+\widetilde{\omega}^{2}\mp4\widetilde{\omega
}\left(  \omega-\frac{\lambda\hbar}{2m}\right)  }-\hbar\left(  \omega
\pm2\widetilde{\omega}\right)  +\nonumber\\
&  \frac{\lambda\hbar^{2}}{2m}\left(  4n\left(  n+l+1\right)  +2l+1\right)
-\left[  \widetilde{\omega}\pm2\left(  \omega+\frac{\lambda\hbar}{2m}\right)
\right]  l \label{87}
\end{align}

The result explicitly shows the contributions of all the terms in eq.$\ref{85}$ and especially those due to the presence of both the spin and the deformation, as well as that of the additional spin-orbit term $2\lambda S_{z}L_{z}$ which can be interpreted as due to the interaction between the two. The same result is found in the case of Dirac oscillator with EUP \cite{Quesne, Sadeghi}.

\section{\textbf{Thermodynamic properties}}

Now we focus on the thermodynamic properties of the system. The partition function at finite temperature $T$ is:

\begin{equation}
Z=\sum\limits_{n=0}^{\infty}e^{-\frac{E_{n}}{k_{B}T}}=\sum\limits_{n=0}
^{\infty}\exp\left(  -\frac{mc^{2}}{k_{B}T}\sqrt{a_{1}+a_{2}n+a_{3}n^{2}
}\right)  \label{88}
\end{equation}
Here $k_{B}$ is the Boltzmann constant and the expressions of the other parameters are:
\begin{align}
a_{1}  &  =1+\frac{2\hbar}{mc^{2}}\left(  \left(  l+1\right)  \sqrt{\left(
\omega-\frac{\lambda\hbar}{2m}\right)  ^{2}+\widetilde{\omega}^{2}}
+\frac{\lambda\hbar}{2m}\left(  2l+1\right)  -\widetilde{\omega}
l-\omega\right) \nonumber\\
a_{2}  &  =\frac{4\hbar}{mc^{2}}\left(  \sqrt{\left(  \omega-\frac
{\lambda\hbar}{2m}\right)  ^{2}+\widetilde{\omega}^{2}}+\frac{\lambda\hbar}
{m}\left(  l+1\right)  \right)  \text{ and }a_{3}=\frac{4\lambda\hbar^{2}
}{m^{2}c^{2}} \label{89}
\end{align}
In order to evaluate this function, we use the Euler-MacLaurin formula:
\begin{equation}
\sum\limits_{n=0}^{\infty}f\left(  n\right)  =\frac{1}{2}f\left(  0\right)
+\int f\left(  x\right)  dx-\sum_{p=1}^{\infty}\frac{1}{\left(  2p\right)
!}B_{2p}f^{\left(  2p-1\right)  }\left(  0\right)  \label{90}%
\end{equation}
$B_{2p}$ are the Bernoulli numbers, $f^{\left(2p-1\right)}$ is the derivative of order $\left(2p-1\right)$ and the integral term is given by:
\begin{align}
I  &  =\frac{2a_{1}}{\sqrt{a_{2}^{2}-4a_{1}a_{3}}}\sum_{n=0}^{\infty}\left(
-1\right)  ^{n}\frac{\left(  2n-1\right)  !!}{\left(  2n\right)  !!}\left(
\frac{4a_{1}a_{3}}{a_{2}^{2}-4a_{1}a_{3}}\right)  ^{n}\nonumber\\
&  \times\left[  \frac{\Gamma\left(  2n+2\right)  }{\chi^{2n+2}}
-\frac{e^{-\chi}}{\left(  2n+2\right)  }\Phi\left(  1,2n+2,\chi\right)
\right]  \label{91}
\end{align}
where we have used $\chi=\frac{mc^{2}}{k_{B}T}\sqrt{a_{1}}$, the new variable $y=\sqrt{1+\frac{a_{2}}{a_{1}}n+\frac{a_{3}}{a_{1}}n^{2}}$and the power series of the square root of the integral:
\begin{equation}
I=\frac{2a_{1}}{\sqrt{a_{2}^{2}-4a_{1}a_{3}}}\int_{1}^{+\infty}\exp\left(
-\chi y\right)\left(1+\frac{4a_{1}a_{3}}{a_{2}^{2}-4a_{1}a_{3}} y^{2}\right)ydy \label{92}
\end{equation}
At high temperatures, we can ignore the first and the third terms in eq.$\ref{90}$ and keep only the integral. Similarly, we neglect the $e^{-\chi}$ term beside the $\chi^{-(2n+2)}$ one in eq.$\ref{91}$. Therefore, the partition function becomes:
\begin{equation}
Z=\left(  \frac{k_{B}T}{mc^{2}}\right)  ^{2}\frac{2}{\sqrt{a_{2}^{2}-4a_{1}a_{3}}} \sum\limits_{n=0}^{\infty}(-1)^{n}\frac{(2n-1)!!}{(2n)!!}\Gamma(2n+2)\sigma^{n} \label{93}
\end{equation}
with:
\begin{equation}
\sigma=\left(  \frac{k_{B}T}{mc^{2}}\right)  ^{2}\frac{4a_{3}}{\left(a_{2}^{2}-4a_{1}a_{3}\right)  } \label{94}
\end{equation}
We restrict ourselves to the first order in $\lambda$, so we can reduce $Z$ to the following simplified form:
\begin{equation}
Z\simeq\frac{\left(  k_{B}T\right)  ^{2}}{2\hbar mc^{2}\sqrt{\widetilde
{\omega}^{2}+\omega^{2}}}-\frac{3\left(  k_{B}T\right)  ^{4}\lambda}{2\hbar
m^{3}c^{4}\sqrt[3]{\widetilde{\omega}^{2}+\omega^{2}}}\left(  1-\frac
{mc^{2}\left(  2\hbar\widetilde{\omega}l+mc^{2}+\hbar\omega\right)  }{6\left(
k_{B}T\right)  ^{2}}\right)  \label{95}
\end{equation}
We neglect the $\left(k_{B}T\right)^{-2}$ term in parentheses and we obtain the high-temperature expansion of the partition function:
\begin{equation}
Z\simeq\frac{\left(  k_{B}T\right)  ^{2}}{2\hbar mc^{2}\sqrt{\widetilde
{\omega}^{2}+\omega^{2}}}\left(  1-\frac{3\left(  k_{B}T\right)  ^{2}\lambda
}{m^{2}c^{2}\left(  \widetilde{\omega}^{2}+\omega^{2}\right)  }\right)
\label{96}
\end{equation}
The first term is the usual partition function for a 2D scalar bosonic oscillator with a uniform magnetic field in the $r$--representation. The second term expresses the contribution that comes from the space deformation through the AdS algebra.

At this stage, we can evaluate all the thermodynamic properties of our system (free energy $F$, mean energy $U$, specific heat $C$ and entropy $S$) using their definitions \cite{Pacheco}:
\begin{equation}
F=-k_{B}T\ln Z\text{,\thinspace}U=k_{B}T^{2}\frac{\partial\ln Z}{\partial T}\text{,\thinspace}C=\frac{\partial U}{\partial T}\text{ and } \,S=-\frac{\partial F}{\partial T} \label{97}
\end{equation}
\begin{gather}
F=-k_{B}T\ln\left(  \frac{\left(  k_{B}T\right)  ^{2}}{2\hbar mc^{2}
\sqrt{\widetilde{\omega}^{2}+\omega^{2}}}\left(  1-\theta\left(k_{B}T\right)  ^{2}\right)  \right) \label{98}\\
U=4k_{B}T\left[  1-\frac{1}{2\left(  1-\theta\left(  k_{B}T\right)^{2}\right)  }\right] \label{99}\\
C=4k_{B}\left[  1-\frac{1+\theta\left(  k_{B}T\right)  ^{2}}{2\left(1-\theta\left(  k_{B}T\right)  ^{2}\right)  ^{2}}\right] \label{100}\\
S=k_{B}\left[  \frac{2-4\theta\left(  k_{B}T\right)  ^{2}}{1-\theta\left(k_{B}T\right)  ^{2}}+\ln\left(  \frac{\left(  k_{B}T\right)^{2}}{2\hbar mc^{2}\sqrt{\widetilde{\omega}^{2}+\omega^{2}}}\left(1-\theta\left(k_{B}T\right)^{2}\right)  \right)\right]  \label{101}
\end{gather}
where we have used the parameter $\theta=\frac{3\lambda}{m^{2}c^{2}\left(\widetilde{\omega}^{2}+\omega^{2}\right)  }$.

We can easily test these expressions in different ways. Using the limit $\widetilde{\omega }\rightarrow0$ (i.e. $B\rightarrow0$), we obtain the thermodynamic results of the deformed 2D scalar bosonic oscillator with AdS commutation relations. If we use the limit $\lambda\rightarrow0$, we obtain the thermal properties of the ordinary 2D bosonic oscillator for both KG and scalar DKP particles in a uniform magnetic field.

We show in figures 1 to 4, the dependence of these thermodynamic properties with the temperature $T$ for different values of the deformation parameter $\lambda$. We chose $\omega=B=1$ and we use the Hartree atomic units ($\hbar=c=k_{B}=m_{e}=1$).

The first remark that appears is that we have a critical point at $T_{c}=\sqrt{\frac{1+\alpha/4}{3\lambda}}=1/\sqrt{3\lambda}$ ($\alpha$ is the fine structure constant) given the presence of the factor $1-\theta\left(
k_{B}T\right)  ^{2}$ in the denominator of $U$, $C$, and $S$ (in eqs.$\ref{99}$, $\ref{100}$ and $\ref{101}$). This factor changes the behavior of these thermodynamic properties as we will see in their graphical representations.

For the free energy, Fig.1 shows that it grows rapidly in the deformed case ($\lambda\neq0$) unlike the normal case where it continues its decrease to infinity; this after a small increase followed by a decrease common to both cases. For a fixed value of $T$ , the free energy increases with the deformation parameter $\lambda$.

Fig.2 shows that there is a discontinuity in the mean energy $U$ at $T_{c}$\ where it decreases rapidly in the vicinity of $T_{c}$ and then it grows slightly as in the ordinary case.

In Fig.3, we see that the specific heat $C$ decreases until the discontinuity point $T_{c}$, then it increases. On the other hand, it is constant for $\lambda=0$ (namely, $C=2k_{B}$).

For the entropy function $S$, we see in Fig.4 that it grows to a maximum value then decreases to infinity and this contrasts with the ordinary case where it has a continuous growth with $T$.

\begin{figure}
\centering
\includegraphics[width=0.5\textwidth]{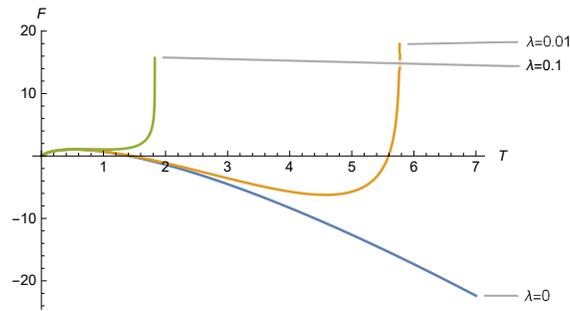}
\caption{The free energy function $F$ according to $T$ for various
values of the deformation parameter}
\label{Fig1}
\end{figure}

\begin{figure}
\centering
\includegraphics[width=0.5\textwidth]{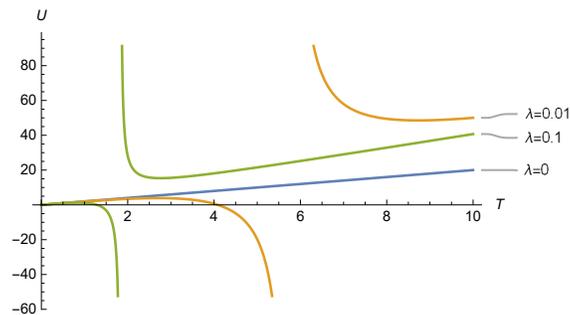}
\caption{The mean energy $U$ according to $T$ for various values
of the deformation values.}
\label{Fig2}
\end{figure}

\begin{figure}
\centering
\includegraphics[width=0.5\textwidth]{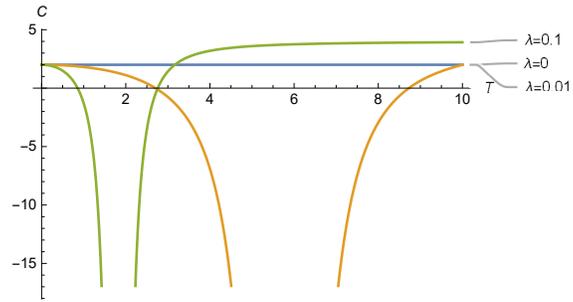}
\caption{The specific heat function $C$ according to $T$ for various values
of the deformation values.}
\label{Fig3}
\end{figure}

\begin{figure}
\centering
\includegraphics[width=0.5\textwidth]{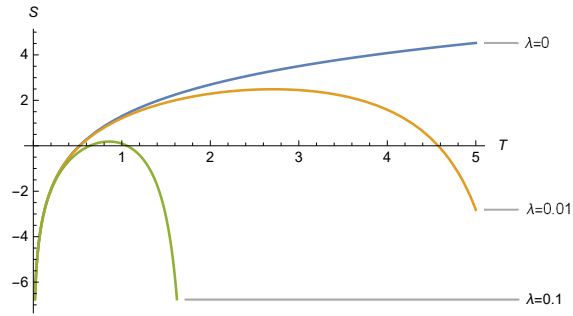}
\caption{The entropy $S$ according to $T$ for various values
of the deformation values.}
\label{Fig4}
\end{figure}

\section{\textbf{Conclusion}}

In this paper, we have studied the exact solutions of the 2D oscillator with an external uniform magnetic field for both KG and scalar DKP equations in the framework of deformed quantum mechanics with Anti--deSitter commutation relations. This AdS deformations lead to a non-zero minimal uncertainty in the measurement of the momentum. We have used the Nikiforov-Uvarov method and so, we obtained the analytical expressions of the bound state energies and the wave functions of the system. In both cases, we expressed analytically eigenfunctions of the system in terms of the Jacobi polynomials and the corresponding eigenenergies with additional corrections depending on the deformation parameter $\lambda$. Our results show that the deformed spectrum remains discrete even for large values of the principal quantum number, whereas, in the ordinary case (without deformations), this spectrum behaves as a continuous one for these values.

In the case of the vector DKP oscillator, since it was almost impossible to find an exact solution to the problem, we limited our study to the non-relativistic limit of this case. The non-zero value of the spin in this case, induced the presence of corrective terms adding to those coming from the deformation. Thus, we were able to obtain the corrective terms representing the effects of the spin and its interactions with the magnetic field, with the orbital moment (the usual spin-orbit term) and a new "spin-orbit" term which represents the interaction of the spin with the orbital moment and the space deformation at the same time. So the spin-orbit contribution is present in the spectrum even in the absence of deformations, but it is enhanced by the influence of the spatial deformation on the system.

Finally, we studied the effects of the space deformation on the thermodynamic properties of our system in the high temperature regime. The graphical representations of our results have showed a clear change in the behavior of these properties. This difference is more pronounced for the free energy which increases in the deformed case while it decreases in the ordinary case and for the entropy which decreases when the deformation is present whereas it increases when there is no more deformation.

\section*{Acknowledgment}

This work was done with funding from the DGRSDT of the Ministry of Higher Education and Scientific Research in Algeria as part of the PRFU B00L02UN070120190003.

\end{document}